\documentstyle[aps,preprint,prl,epsfig]{revtex}
\begin{document}  
\title{Anharmonic double-$\gamma$ vibrations in nuclei
and their description in the interacting boson model}
\author{
J.E.~Garc\'{\i}a--Ramos$^1$,
C.E.~Alonso$^1$, 
J.M.~Arias$^1$, and
P.~Van Isacker$^2$} 
\address{
$^1$Departamento de F\'{\i}sica At\'omica, Molecular y Nuclear,
Universidad de Sevilla, Apartado 1065, 41080 Sevilla, Spain} 
\address{
$^2$Grand Acc\'el\'erateur National d'Ions Lourds,
B.P.~5027, F-14076 Caen Cedex 5, France}
\date{\today}
\maketitle
          
\begin{abstract}
Double-$\gamma$ vibrations in deformed nuclei
are studied in the context of the interacting boson model
with special reference to their anharmonic character.
It is shown that large anharmonicities
can be obtained with interactions
that are (at least) of three-body nature between the bosons.
As an example
the $\gamma$ vibrations of the nucleus
$^{166}_{\phantom{0}68}$Er$^{\phantom{00}}_{98}$
are studied in detail.
\end{abstract}

\pacs{PACS numbers: 21.60.-n, 21.60.Fw, 21.60.Ev}

\draft

Nuclear quadrupole shape oscillations can be of two types:
$\beta$ or $\gamma$ vibrations \cite{BM75}.
The $\beta$ vibration preserves axial symmetry
and a one-quantum excitation gives rise to a $K=0$ band
where $K$ is the projection of the angular momentum
on the axis of symmetry of the nucleus.
A $\gamma$ vibration breaks axial symmetry and leads to a $K=2$ band.
Although their existence has been conjectured a long time ago \cite{BM53},
the observation and interpretation of $\beta$-vibrational $K^\pi=0^+$ bands
is still fraught with questions and difficulties.
In contrast, $\gamma$-vibrational $K^\pi=2^+$ bands
are systematically observed in deformed nuclei
and their properties are correspondingly better understood.

Since single-$\gamma$ vibrations are so well established,
it is natural to search for double-$\gamma$ vibrations
and to examine their harmonic nature
(i.e., whether they occur at twice the energy of the single vibration.)
Two intrinsic $K=2$ quanta can be combined parallel or antiparallel
and hence lead to two bands:
one with $K=0$ and another with $K=4$.
The experimental identification of double-$\gamma$ vibrations
in deformed nuclei is difficult
since they are expected to lie above the pairing gap
and to mix with two-quasiparticle excitations,
resulting in fragmentation and
a corresponding reduction in the collectivity of the states.
During the last few years, however,
a steady improvement of experimental techniques
has allowed the measurement of low-spin states
in the energy region of interest \cite{Fahl88,Born93,Belg96}.
This possibility has reopened the old debate
on the existence of two-phonon ($\beta$ or $\gamma$) vibrational states
and their properties.
Experiments have been reported recently
pointing out the existence
of double-$\gamma$ vibrations in several deformed nuclei
with a wide range of anharmonicities
\cite{Born91,Fahl96,Garr97,Gues95,Apra94,Corm97}.
In particular, in Refs.~\cite{Fahl96,Garr97}
the first observation of the $K^\pi=0^+$ {\em and} $K^\pi=4^+$
double-$\gamma$ states in one nucleus, $^{166}$Er, is reported. 
They are observed at 1.949 MeV and 2.029 MeV, respectively.
This information is of great interest
since it provides a stringent test of nuclear models;
for instance, the Quasiphonon Nuclear Model (QPNM)
predicts no $K^\pi=0^+$ two-phonon state
below 2.5 MeV in $^{166}$Er \cite{Solo95}.
Several calculations of two-phonon states,
using either phenomenological or microscopic models,
are available,
particularly for $^{166}$Er and $^{168}$Er
\cite{Solo95,BM82,Dumi82,Jamm88,Yosh86,Solo94}.
One of the models employed is the Interacting Boson Model (IBM) \cite{Iach87}.
In the simplest version of this model, referred to as IBM-1, 
an even-even nucleus with $n$ valence nucleons
is treated as a system of $N={\displaystyle{n\over2}}$ bosons
with $\ell=0$ ($s$ bosons) or $\ell=2$ ($d$ bosons).
In the usual formulation of the model
only up to two-body interactions between the bosons are taken.

What are the predictions of IBM
with regard to two-phonon states in deformed nuclei
and their (an)harmonic nature?
It was pointed out some time ago by Bohr and Mottelson \cite{BM82}
that the IBM-1 is unable to accommodate large anharmonicities,
as observed for instance in $^{168}$Er.
Subsequently, it was shown that these can be described
but require $g$ bosons with $\ell=4$
in addition to the $s$ and $d$ bosons ($sdg$-IBM) \cite{Yosh86}.
More recently, we reported a study of two-phonon states in IBM-1
treated in the intrinsic frame \cite{Garc98} 
and showed that the IBM-1 is a harmonic model
in the limit of large boson number.
Anharmonicities can only exist for finite boson number
and they are always small
if only up to two-body interactions are considered.
It was also suggested that anharmonicity in the model
is linked to triaxiality.
Since it is known that IBM-1 with only up to two-body interactions
cannot give rise to a stable triaxial minimum,
the model's capability for describing anharmonicities
depends on the inclusion in the Hamiltonian of higher-order interactions,
some of which are known to induce triaxial shapes \cite{Isac81,Heyd84}.

In this article the relation between three-body interactions in IBM-1
and the anharmonicity of $\gamma$ vibrations in deformed nuclei
is investigated.
Although the analysis presented is not exhaustive,
it is shown that anharmonic behavior
can be obtained with reasonable three-body interactions.
As an example, the energy and E2 transition properties
of the $\gamma$ vibrations of the nucleus $^{166}$Er
are studied in detail.
In addition, the nature of the $0^+_2$ state in the same nucleus,
which has been the subject of an intense debate
in the last few years \cite{Garr2-97,Cast94,Burk95,Gunt96},
is briefly discussed.

The Hamiltonian adopted in the following
includes a quadrupole-quadrupole term,
a rotational $\hat L^2$ term,
and three-body interactions between the $d$ bosons,
\begin{equation}
\hat H=
-\kappa\hat Q\cdot\hat Q+ 
\kappa'\hat L \cdot \hat L+
\sum_{k l}\theta_l
\left((d^\dagger\times d^\dagger)^{(k)}\times d^\dagger\right)^{(l)}
\cdot
\left((\tilde d\times\tilde d)^{(k)}\times\tilde d\right)^{(l)},
\label{ham}
\end{equation}
where $\cdot$ denotes scalar product,
$\tilde d_\mu=(-1)^\mu d_{-\mu}$,
$\hat Q$ is the boson quadrupole operator,
and $\hat L$ is the angular momentum operator:
\begin{equation}
\hat Q=s^{\dagger}\tilde d+d^\dagger\tilde s
+\chi\left(d^\dagger\times\tilde d\right)^{(2)},
\qquad
\hat L=\sqrt{10} \left(d^\dagger\times\tilde d\right)^{(1)}.
\label{ql}
\end{equation}
Five independent three-body $d$-boson interactions exist
which have $l=0$, 2, 3, 4, and 6.
Interactions with the same $l$ but different $k$
are not independent
but differ by a normalization factor only \cite{Isac81}.
The combinations
$(k,l)=(2,0)$, (0,2), (2,3), (2,4), and (4,6) are chosen here.

The Hamiltonian (\ref{ham}) is certainly not the most general
that can be considered.
Notably, a vibrational term $\epsilon_d\hat n_d$
which dominates in spherical nuclei is omitted
since it is thought of lesser importance
in the deformed nuclei considered here.
It is clear that the inclusion of such additional terms
might improve the quality of detailed fits to particular nuclei
such as the one for $^{166}$Er presented below.
Finally, of all possible three-body interactions
only those between the $d$ bosons are retained here
since these are most crucial
for obtaining a stable triaxial minimum \cite{Isac81}.

For the discussion of anharmonicities of $\gamma$ vibrations
it is convenient to define the following ratios of excitation energies:
\begin{equation}
R_0^\gamma\equiv
{{E_{\rm x}(0^+_{\gamma\gamma})}
\over
{E_{\rm x}(2^+_\gamma)-E_{\rm x}(2^+_1)}},
\qquad
R_4^\gamma\equiv
{{E_{\rm x}(4^+_{\gamma\gamma})-E_{\rm x}(4^+_1)}
\over
{E_{\rm x}(2^+_\gamma)-E_{\rm x}(2^+_1)}},
\label{ratios}
\end{equation}
where $0^+_{\gamma\gamma}$ and $4^+_{\gamma\gamma}$
are the band heads of the $K^\pi=0^+$ and $K^\pi=4^+$ double-$\gamma$ bands,
respectively.
It should be noted that the quantities $R_K^\gamma$
do not depend upon the $\hat L^2$ term in the Hamiltonian;
if a single three-body term is included
they depend on two parameters, $\chi$ and the ratio $\theta_l/\kappa$.
In the present work the identification of the states
$0^+_{\gamma\gamma}$ and $4^+_{\gamma\gamma}$ is based on the B(E2)
values for decaying to the single gamma state.
In Fig.~\ref{fig-chi} the quantities $R_K^\gamma$
are plotted (for $N=15$ bosons)
as a function of the quadrupole parameter $\chi$
(varying between its SU(3) and O(6) values $-{1\over2}\sqrt7$ and 0)
in the absence of three-body interactions.
The ratio $R_4^\gamma$
remains about constant and of the order 1.8;
$R_0^\gamma$ shoots up for small $|\chi|$.
Close to the O(6) limit the concept of a $\gamma$ vibration
is not well defined and so nothing is plotted for $|\chi|<0.15$.
The value of $\chi$ is constrained by E2 transition probabilities
and in deformed rare-earth nuclei it ranges typically
between $-0.4$ and $-0.7$ \cite{Cast88}.
From Fig.~\ref{fig-chi} it is clear that 
no substantial anharmonicity occurs
in the $\gamma$ vibration for these values of $\chi$.

In Fig.~\ref{fig-theta} the influence of the various three-body interactions
is shown for a typical value of $\chi$ ($\chi=-0.5$)
and for $N=15$ bosons.
It is seen that
$\gamma$-vibrational anharmonic behavior is obtained
which can be different for the $K^\pi=0^+$ and $K^\pi=4^+$ bands
(e.g., positive for the former while negative for the latter.)
Care has been taken to plot results only up to values of $\theta_l$
that do not drastically alter the character of rotational spectrum;
beyond these values, the three-body interaction,
being of highest order in the Hamiltonian~(\ref{ham}),
becomes dominant.
Also shown in Fig.~\ref{fig-theta} are the ratios $R_K^\gamma$
as observed in $^{166}$Er \cite{Fahl96,Garr97},
$R_0^\gamma=2.76$ and $R_4^\gamma=2.50$.
This simple analysis shows that,
on purely phenomenological grounds,
the appropriate three-body interaction
with the correct anharmonic character
for the $K^\pi=0^+$ and $K^\pi=4^+$ bands in $^{166}$Er,
has $l=4$.

Figure~\ref{fig-spec}
shows the experimental spectrum of $^{166}$Er \cite{Fahl96,Garr97}
and compares it to the eigenspectrum of Hamiltonian (\ref{ham})
with an $l=4$ three-body interaction.
The parameters are $\kappa=23.8$ keV, $\chi=-0.55$,
$\kappa'=-1.9$ keV, and $\theta_4=93.9$ keV,
with boson number $N=15$.
With these values the
calculated excitation energies of the double-$\gamma$ band heads
are 1926 keV and 1972 keV for the $K^\pi=0^+$ and $K^\pi=4^+$ levels,
respectively, leading to the ratios
$R_0^\gamma=2.82$ and $R_4^\gamma=2.45$,
in excellent agreement with observation.
Note, however, that although all $\gamma$-band heads
are well reproduced by the calculation,
problems arise for the moments of inertia,
in particular of the $\gamma$ band.
An extensive survey of combinations of cubic $d$-boson interactions
has shown that it is difficult
to substantially improve upon this fit
although it is of course near-impossible
to do an exhaustive search
of the complex parameter space of all three-body interactions.
In contrast, exploratory searches
with simple quartic Hamiltonians
quickly yield the correct result
with respect to both band-head energies and moments of inertia.

For the calculation of E2 transition probabilities
the Consistent-Q Formalism (CQF) \cite{Warn82} is adopted
by using the E2 transition operator
\begin{equation}
\hat T({\rm E}2)=e_{\rm eff}\;\hat Q,
\label{TE2}
\end{equation}
where $\hat Q$ is the boson quadrupole operator
used in the Hamiltonian (\ref{ham})
and $e_{\rm eff}$ is a boson effective charge,
determined from the observed $B({\rm E}2;2^+_1\rightarrow0^+_1)$ value.
It should be noted that the inclusion of three-body terms in
the Hamiltonian would allow the use of a two-body
E2 operator. However, we have not tried to do that in order to keep
the calculation on the anharmonicity of the double-gamma excitation as
simple as possible.
In table~\ref{tab-trans} the observed $B$(E2) values and ratios
concerning the $\gamma$ vibrational band heads in $^{166}$Er
are summarized and compared to the theoretical results
obtained with $e_{\rm eff}^2=(1.83)^2$ Weisskopf units (W.u.).
A good overall agreement is found
but for the decay of the $0^+_2$ state:
the $B({\rm E}2;0^+_2\rightarrow2^+_\gamma$) value
is overpredicted by more than an order of magnitude
while the $B({\rm E}2;0^+_2\rightarrow2^+_1$) value
is too small by a factor two.
This casts doubt on the interpretation of the $0^+_2$ observed 
at 1460 keV as the $\beta$-band head.
Previous interpretations of this state are contradictory:
it is considered as the $\beta$-band head in \cite{BM75}
but as a two-quasiparticle state in \cite{Garr2-97}
while Casten and von Brentano \cite{Cast94}
claim it is a collective phonon excitation built on the $\gamma$ band.
Other $0^+$ states are found in $^{166}$Er
at slightly higher energy \cite{Garr2-97}
but none has the decay pattern in agreement with the present calculation.
A possible explanation is that collective strength
is fragmented through mixing with two-quasiparticle states
which are absent from the IBM-1 model space.

In summary,
three-body interactions in the IBM-1
can account for a wide variety of $\gamma$-vibrational anharmonicities
in nuclei
such as for instance those observed in $^{166}$Er
but not without substantially changing
the moments of inertia of various bands.
The knowledge of the {\em two} double-$\gamma$ vibrational bands
($K^\pi=0^+$ and $K^\pi=4^+$) in a single nucleus
provides a stringent test of nuclear models
and, specifically, of the type and strength
of three-body interactions in IBM-1.
More experiments on double-$\gamma$ vibrations are thus called for
since they should provide essential information
concerning the systematic behavior of these states
and hence the interactions involved.
From the theoretical side,
a systematic analysis of {\em all} three-body interactions
and not just those between the $d$ bosons
seems in order.
Once a fuller knowledge is acquired of the systematic behavior
of the interactions necessary to reproduce the observed anharmonicities,
one may then attempt an understanding on a microscopic level.

We are grateful to F.~Iachello, A.~Vitturi, and C.~Volpe
for valuable comments.
This work has been supported in part by the Spanish DGICYT under contract
No.~PB98--1111 and by one IN2P3 (France)-CICYT (Spain) agreement.

\begin{figure}
\caption{The ratios $R_K^\gamma$ (as defined in the text)
as a function of $\chi$.
The Hamiltonian (1) is used with $\theta_l=0$;
the boson number is $N=15$.}
\label{fig-chi}
\end{figure}

\begin{figure}
\caption{The ratios $R_K^\gamma$ (as defined in the text)
as a function of $\theta_l/\kappa$ for different $l$.
The Hamiltonian (1) is used with $\chi=-0.5$;
the boson number is $N=15$.
The dashed lines give the experimental values
for the corresponding ratios in $^{166}$Er.}
\label{fig-theta}
\end{figure}

\begin{figure}
\caption{Experimental (a) and calculated (b) spectrum for $^{166}$Er. 
The theoretical results are obtained with the Hamiltonian (1)
with $\kappa=23.8$ keV, $\chi=-0.55$, $\kappa'=-1.9$ keV,
and $\theta_4=93.9$ keV.
The boson number is $N=15$.}
\label{fig-spec}
\end{figure}

\begin{table}
\caption{Observed and calculated $B$(E2) values and ratios for $^{166}$Er.
The E2 operator (4) is used with $e_{\rm eff}^2=(1.83)^2$ W.u.\
and $\chi=-0.55$.}
\begin{tabular}{lcc}
&\multicolumn{2}{c}{$B$(E2) value or ratio}\\
\cline{2-3}
&Observed&Calculated\\
\hline
$B({\rm E}2;2_1^+\rightarrow0_1^+)$ (W.u.)&$214\pm10$ $^a$ &214\\
$B({\rm E}2;4_1^+\rightarrow2_1^+)$ (W.u.)&$311\pm10$ $^a$ &304\\
$B({\rm E}2;2_\gamma^+\rightarrow0_1^+)$ (W.u.)&$5.5\pm0.4$ $^a$ &5.3\\
${\displaystyle
{B({\rm E}2;0_2^+\rightarrow2_1^+)\over
 B({\rm E}2;2_\gamma^+\rightarrow0_1^+)}}$&$0.49\pm0.19$ $^b$&0.21\\

${\displaystyle
{B({\rm E}2;0_2^+\rightarrow2_\gamma^+)\over
 B({\rm E}2;2_\gamma^+\rightarrow0_1^+)}}$&$0.44\pm0.13$ $^b$&6.2\\

${\displaystyle
{B({\rm E}2;0_{\gamma\gamma}^+\rightarrow2_\gamma^+)\over
 B({\rm E}2;2_\gamma^+\rightarrow0_1^+)}}$&$3.8\pm1.3$ $^c$ 
($2.2{{\textstyle +1.1}\atop{\textstyle -0.7}}$ $^d$)&3.2\\

${\displaystyle
{B({\rm E}2;4_{\gamma\gamma}^+\rightarrow2_\gamma^+)\over
 B({\rm E}2;2_\gamma^+\rightarrow0_1^+)}}$&$1.3\pm0.4$ $^c$
($0.9\pm0.3$ $^d$)&2.5\\
\end{tabular}

$^a$ From Ref.~\cite{Shur92}.

$^b$ From Ref.~\cite{Garr2-97}. 

$^c$ From Ref.~\cite{Garr97}. 

$^d$ From Ref.~\cite{Fahl96}.

\label{tab-trans}
\end{table}

\begin{references}

\bibitem{BM75}
A.~Bohr and B.R.~Mottelson,
{\it Nuclear structure, Vol.~II}
(Benjamin, Reading, Massachusetts, 1975).

\bibitem{BM53}
A.~Bohr and B.R.~Mottelson,
Mat.\ Fys.\ Medd.\ Dan.\ Vid.\ Selsk.\ {\bf27}, no 16 (1953).

\bibitem{Fahl88}
C.~Fahlander {\it et al.}, Nucl.\ Phys.\ A {\bf485}, 327 (1988).

\bibitem{Born93}
H.G.~B\"orner and J.~Jolie, J.\ Phys.\ G {\bf19}, 217 (1993).

\bibitem{Belg96}
T.~Belgya, G.~Moln\'ar, and S.W.~Yates, Nucl.\ Phys.\ A {\bf607}, 43 (1996).

\bibitem{Born91}
H.G.~B\"orner, J.~Jolie, S.J.~Robinson, B.~Krusche, R.~Piepenbring, 
R.F.~Casten, A.~Aprahamian, and J.P.~Draayer, Phys.\ Rev.\ Lett.\ 
{\bf66}, 691 (1991).

\bibitem{Fahl96}
C.~Fahlander, A.~Axelsson, M.~Heinebrodt, T.~H\"artlein, and D.~Schwalm,
Phys.\ Lett.\ B {\bf388}, 475 (1996).

\bibitem{Garr97}
P.E.~Garrett, M.~Kadi, M.~Li, C.A.~McGrath, V.~Sorokin, M.~Yeh, and 
S.W.~Yates , Phys.\ Rev.\ Lett.\ {\bf78}, 4545 (1997).

\bibitem{Gues95}
A.~Guessous {\it et al.}, Phys.\ Rev.\ Lett.\ {\bf75}, 2280 (1995).

\bibitem{Apra94}
A.~Aprahamian, X.~Wu, S.M.~Fischer, W.~Reviol, and J.X.~Saladin,
in {\it Proceedings of the 8th International Symposium
on Capture Gamma-Ray Spectroscopy}, edited by J.~Kern
(World Scientific, Singapore, 1994), p.~57.

\bibitem{Corm97}
F.~Corminboeuf, J.~Jolie, H.~Lehmann, K.~Fohl, F.~Hoyler,
H.G.~B\"orner, C.~Doll, P.E.~Garrett, Phys.\ Rev.\ C {\bf56}, 
R1201 (1997).

\bibitem{Solo95}
V.G.~Soloviev, A.V.~Sushkov, and N.~Yu.~Shirikova, Phys.\ Rev.\ C
{\bf51}, 551 (1995).

\bibitem{BM82}
A.~Bohr and B.R.~Mottelson, Phys. Scr. {\bf 25}, 28 (1982).

\bibitem{Dumi82}
T.S.~Dumitrescu and I.~Hamamoto, Nucl.\ Phys.\ A {\bf 383}, 205 (1982).

\bibitem{Jamm88}
M.K.~Jammari and R.~Piepenbring, Nucl.\ Phys.\ A {\bf 487}, 77 (1988).

\bibitem{Yosh86}
N.~Yoshinaga, Y.~Akiyama, and A.~Arima, Phys.\ Rev.\ Lett. {\bf 56},
1116 (1986); Phys.\ Rev.\ C {\bf38}, 419 (1988). 

\bibitem{Solo94}
V.G.~Soloviev, A.V.~Sushkov, and N.~Yu.~Shirikova, J.\ Phys.\ G {\bf20},
113 (1994).

\bibitem{Iach87}
F.~Iachello and A.~Arima, {\it The interacting boson model}
(Cambridge University Press, Cambridge, 1987).

\bibitem{Garc98}
J.E.~Garc\'{\i}a-Ramos, C.E.~Alonso, J.M.~Arias, P.~Van Isacker, and
A.~Vitturi, Nucl.\ Phys.\ A {\bf 637}, 529 (1998).

\bibitem{Isac81}
P.~Van~Isacker and J.Q.~Chen, Phys.\ Rev.\ C {\bf24}, 684 (1981).

\bibitem{Heyd84}
K.~Heyde, P.~Van Isacker, M.~Waroquier, and J.~Moreau,
Phys.\ Rev.\ C {\bf29}, 1420 (1984).

\bibitem{Garr2-97}
P.E.~Garrett, M.~Kadi, C.A.~McGrath, V.~Sorokin, M.~Li, M.~Yeh, 
and S.W.~Yates, Phys.\ Lett.\ B {\bf400}, 250 (1997). 
 
\bibitem{Cast94}
R.F.~Casten and  P.~von Brentano, Phys.\ Rev.\ C {\bf50}, R1280 (1994);
Phys.\ Rev.\ C {\bf51}, 3528 (1995).

\bibitem{Burk95}
D.G.~Burke and P.C.~Sood, Phys.\ Rev.\ C {\bf51}, 3525 (1995).

\bibitem{Gunt96}
C. G\"unther, S.~Boehmsdorff, K.~Freitag, J.~Manns, and U.~Muller,
Phys.\ Rev.\ C {\bf54}, 679 (1996).

\bibitem{Cast88}
R.F.~Casten and D.D.~Warner,
Rev.\ Mod.\ Phys.\ {\bf60}, 389 (1988).

\bibitem{Warn82}
D.D.~Warner and R.F.~Casten, Phys.\ Rev.\ Lett.\ {\bf48}, 1385 (1982).

\bibitem{Shur92}
E.N.~Shurshikov and  N.V.~Timofeeva, Nucl.\ Dat.\ Sheets {\bf 67}, 45 
(1992).

\end{references}
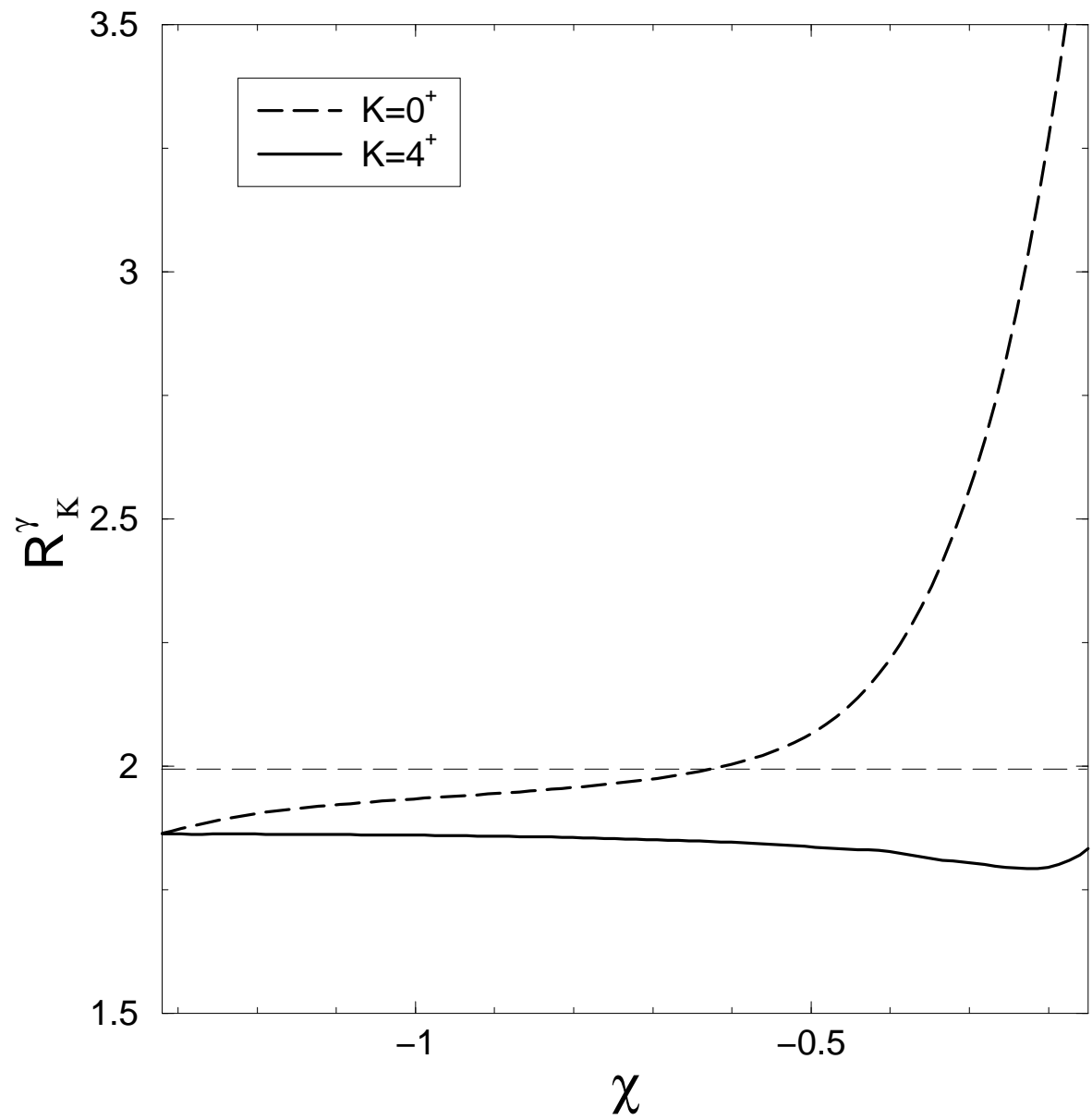
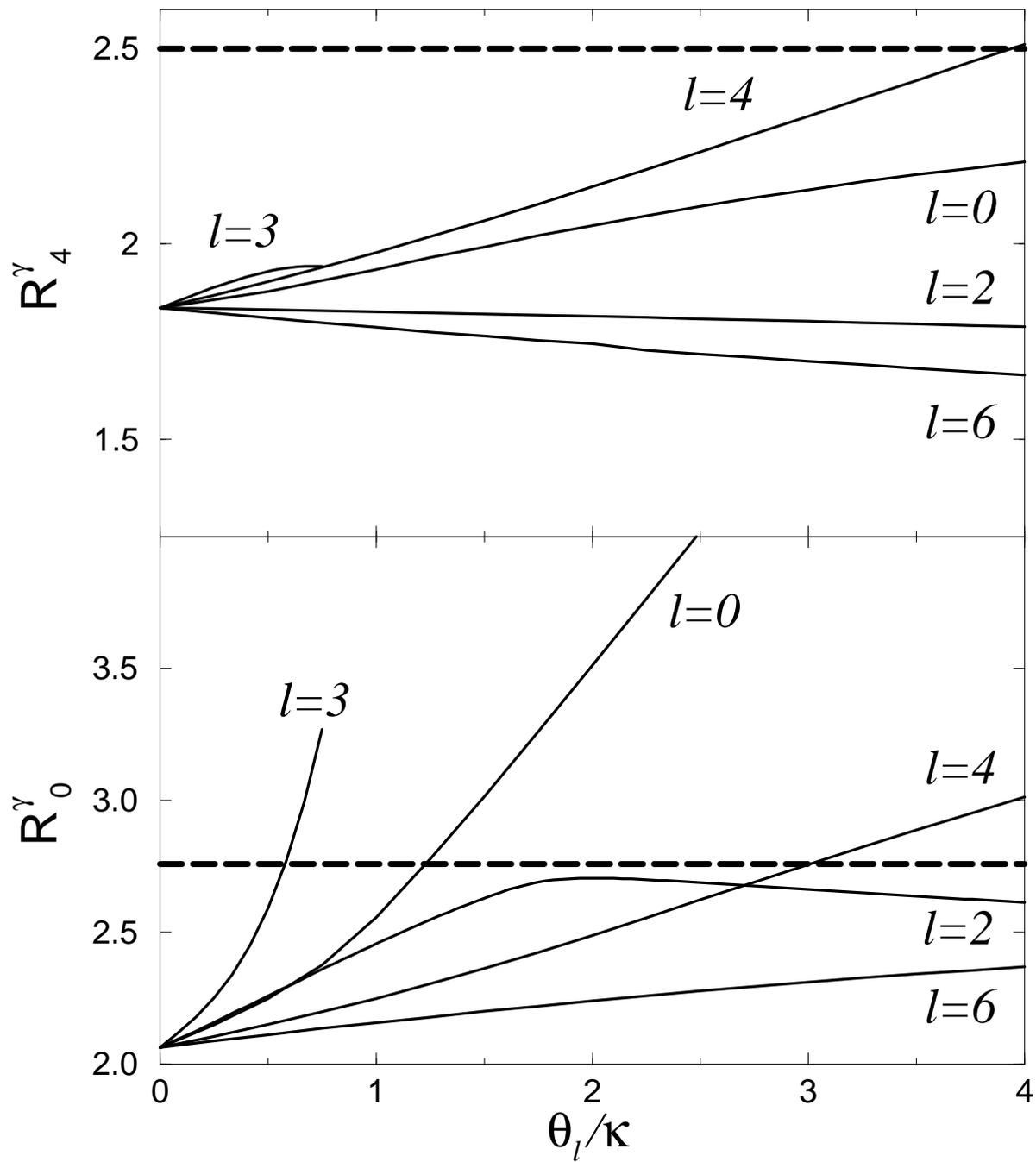
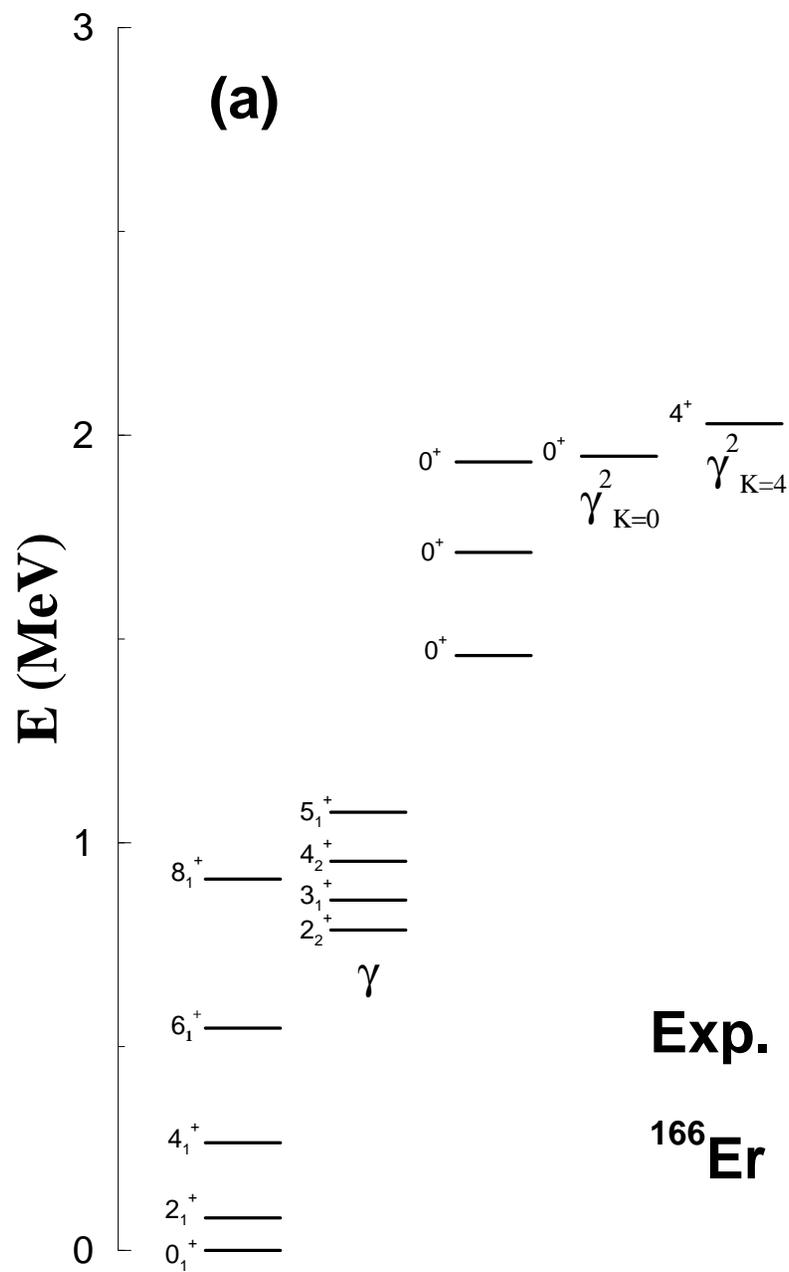
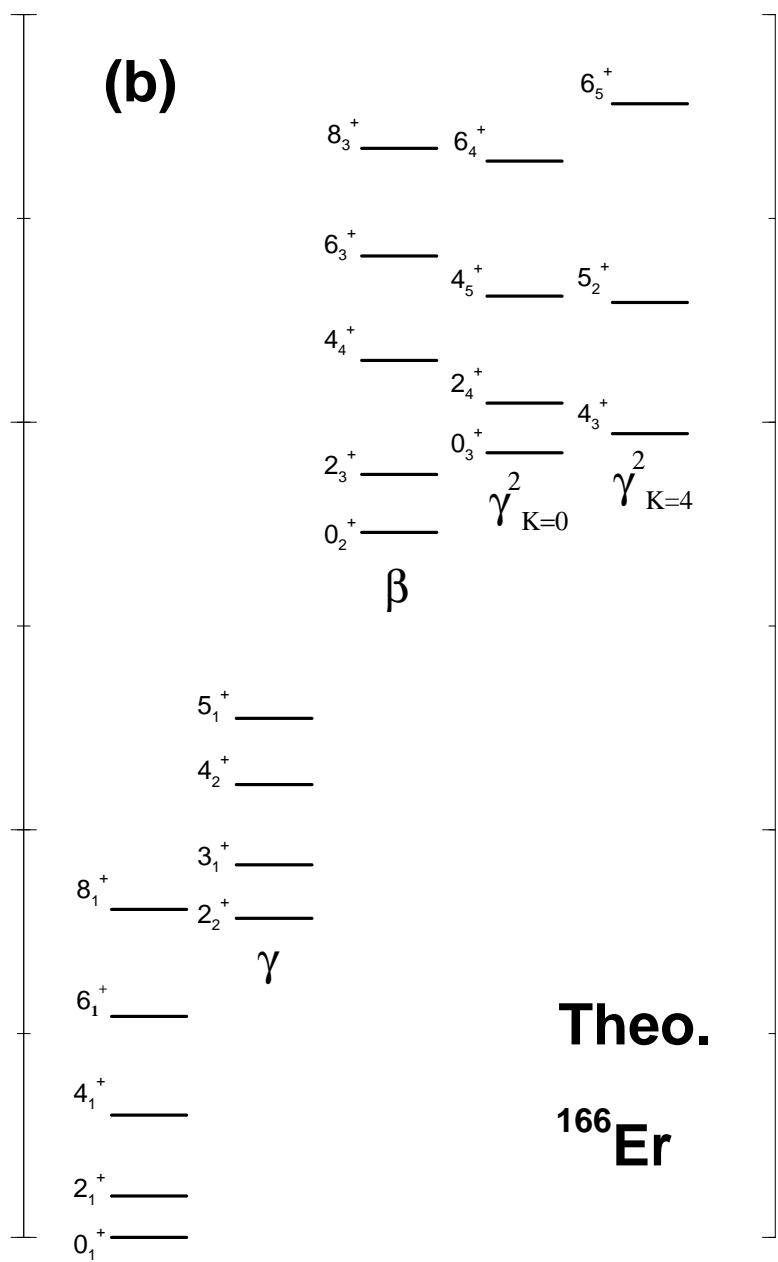
\end{document}